\documentclass[twocolumn,trackchanges, times]{aastex631}

\usepackage{CJK}

\newcommand{\sevenrm}{\rm\scriptsize}

\def\Oiii{[O{\sevenrm\,III}]}
\def\Nii{[N{\sevenrm\,II}]}
\def\Sii{[S{\sevenrm\,II}]}

\def\Mgii{Mg{\sevenrm\,II}}
\def\Feii{Fe{\sevenrm\,II}}
\def\Civ{C{\sevenrm\,IV}}
\def\Siiv{Si{\sevenrm\,IV}}

\def\Heii{He\,{\sc II}}

\def\Cii{C{\sevenrm\,II}}
\def\Ciii{C{\sevenrm\,III}}
\def\Feii{Fe{\sevenrm\,II}}

\received{Dec 4, 2024}
\revised{Feb 13, 2025}
\accepted{Feb 13, 2025}

\submitjournal{ApJL}

\begin{document}

\title{\large Fading Light, Fierce Winds: JWST Snapshot of a Sub-Eddington Quasar at Cosmic Dawn}


\author[0000-0002-6221-1829]{Jianwei Lyu \begin{CJK}{UTF8}{gbsn}(吕建伟)\end{CJK}}
\affiliation{Steward Observatory, University of Arizona,
933 North Cherry Avenue, Tucson, AZ 85721, USA}

\author[0000-0003-2303-6519]{George H. Rieke}
\affiliation{Steward Observatory, University of Arizona,
933 North Cherry Avenue, Tucson, AZ 85721, USA}

\author[0000-0002-9720-3255]{Meredith Stone}
\affiliation{Steward Observatory, University of Arizona, 933 North Cherry Avenue, Tucson, AZ 85721, USA}

\author[0000-0002-9288-9235]{Jane Morrison}
\affiliation{Steward Observatory, University of Arizona, 933 North Cherry Avenue, Tucson, AZ 85721, USA}

\author[0000-0002-8909-8782]{Stacey Alberts}
\affiliation{Steward Observatory, University of Arizona,
933 North Cherry Avenue, Tucson, AZ 85721, USA}

\author[0000-0002-5768-738X]{Xiangyu Jin}
\affiliation{Steward Observatory, University of Arizona, 933 North Cherry Avenue, Tucson, AZ 85721, USA}

\author[0000-0003-3307-7525]{Yongda Zhu}
\affiliation{Steward Observatory, University of Arizona, 933 North Cherry Avenue, Tucson, AZ 85721, USA}

\author[0000-0003-3762-7344]{Weizhe Liu \begin{CJK}{UTF8}{gbsn}(刘伟哲)\end{CJK}}
\affiliation{Steward Observatory, University of Arizona, 933 North Cherry Avenue, Tucson, AZ 85721, USA}

\author[0000-0001-5287-4242]{Jinyi Yang}
\affiliation{Department of Astronomy, University of Michigan, 1085 S. University Ave., Ann Arbor, MI 48109, USA}


\begin{abstract}
The majority of most luminous quasars during the epoch of reionization accrete
near or above the Eddington limit, marking the vigorous growth of primitive
supermassive black holes (SMBHs). However, their subsequent evolution and
environmental impact remain poorly characterized. We present JWST/NIRSpec prism
IFU observations of HSC J2239+0207, a low-luminosity quasar at $z\sim6.25$
likely in a late stage of mass assembly with an overmassive SMBH relative to its
host galaxy. Using H$\beta$ and H$\alpha$ broad emission lines, we estimate an
SMBH mass $M_{\rm BH}\sim3\times10^8~M_{\odot}$ and confirm its sub-Eddington
accretion at $\lambda_{\rm Edd}\sim0.4$. Strong \Feii ~emission and a proximity
zone of typical size suggest a metal-rich, highly evolved system. In the far-UV,
this quasar presents strong broad-absorption-line features, indicative of
high-velocity winds ($\nu\sim10^4~{\rm km/s}$). Meanwhile, minimal dust
reddening is inferred from the quasar continuum and broad-line Balmer decrement,
suggesting little dust along the polar direction. Most interestingly, we
identify a gas companion $\sim$5 kpc from the quasar with a high \Oiii/H$\beta$
ratio ($\gtrsim10$), likely representing outflowing gas blown away by AGN
feedback. These results highlight HSC J2239+0207 as a likely fading quasar in
transition, providing rare insights into SMBH evolution, AGN feedback, and
AGN-galaxy interactions in the early Universe.

\end{abstract}


\section{Introduction} \label{sec:intro}

In the local Universe, correlations between the masses of supermassive
black holes (SMBHs) and the properties of their host galaxies have been well
established, postulating a possible co-evolution of SMBH and galaxy \citep[e.g.,
see review by][]{Kormendy2013}. Over the past two decades, characterizing and
interpreting the relation between SMBHs and their hosts across cosmic time
have been a key mission in extragalactic astronomy \citep[e.g.,][]{Alexander2012,
Heckman2014, Somerville2015}. The situation in the epoch of
reionization ($z\sim6$) has attracted  particular interest. Observations
of luminous quasars at these redshifts reveal SMBHs with masses exceeding
10$^9~M_\odot$, providing a unique window into the early growth of SMBHs and the
build-up of their host galaxies when the Universe was less than 1 Gyr old \citep[e.g., see a recent review by][]{Fan2023}.

With unprecedented sensitivity, spatial resolution, and spectral coverage at
near- to mid-IR wavelengths, JWST follow-up observations of these bright quasars
have provided valuable insights into the quasar evolution, host galaxy
properties and large-scale environment during the reionization epoch
\citep[e.g.,][]{Ding2023, Yang2023, Wang2023, Stone2024}. These most luminous
quasars are found to be accreting at or above the  Eddington limit
\citep[e.g.,][]{Fan2023} and the AGNs are metal-rich \citep[e.g.,][]{JiangD2024}
with identical spectral properties to their low-$z$ counterparts
\citep[e.g.,][]{Yang2023}. Compared to the masses of the SMBHs, the host
galaxies are typically undermassive \citep[e.g.,][]{Stone2024, Yue2024} and reside in
overdense environments (e.g., \citealt{Wang2023}; \citealt{Eilers2024}). In
theory, with the consumption of the surrounding materials, these quasars are
expected to fade away with time and leave some signatures of the AGN feedback in
the environment {\citep[e.g.,][]{Volonteri2010, Inayoshi2022}}. Nevertheless, such phenomena are poorly
characterized by observations.

In this paper, we present the JWST/NIRSpec IFU observations of HSC J2239+0207,
an intriguing quasar at $z\sim6.25$ discovered by the Subaru High-$z$
Exploration of Low-Luminosity Quasars project \citep{Izumi2019}. Compared to
other bright quasars at $z\gtrsim6$, the luminosity of J2239+0207 is relatively
low ($M_{\rm 1450, optical}=-24.7$ or $L_{\rm bol}\sim2.5\times10^{12}~L_\odot$)
and near the break of the quasar luminosity function at $z\sim6$
\citep{Matsuoka2018}, making it more representative of the general quasar
population during the reionization epoch. Ground-based spectral follow-up of
this object put a tight constraint on the object redshift at $z$=6.2498 and a
black hole mass estimate of $M_{\rm BH}\sim(0.6$ --$ 1)\times10^9 M_\odot$
(based on MgII and CIV line widths; \citealt{Onoue2019}) with a corresponding
Eddington ratio of $\sim$20\%. Based on careful PSF subtractions of multi-band
JWST/NIRCam medium band images of this quasar, \cite{Stone2023} reported the
detection of the host galaxy stellar output  and estimated a total stellar mass
of $\sim10^{10}$~$M_\odot$. Combined with the SMBH mass measurement, this yields
a black hole to galaxy stellar mass ratio $\sim$15 times larger than predicted
by the local $M_{\rm BH}$-$M_*$ relation. The combination of the low Eddington
ratio and over-massive nature of the SMBH indicates that HSC J2239+0207 is
likely in a late stage of  mass assembly. That is,  as its accretion decreases and it gradually fades, its host galaxy is expected to grow continuously so the system will eventually approach more closely the typical $M_{\rm BH}$-$M_*$ relation. Therefore, it offers us a unique opportunity to study SMBH evolution and AGN-galaxy interaction
at cosmic dawn.

We organize this paper as follows: Section~\ref{sec:data} describes the
NIRSpec/IFU data reduction and the methods used for spectral analysis. In
Section~\ref{sec:results}, we provide various constraints on the quasar
properties, including e.g., the black hole mass, quasar proximity zone, iron
emission strength, broad absorption lines, dust attenuation.
Section~\ref{sec:comp} reports the discovery of a gas companion near the quasar
redshift and characterizes its properties. Section~\ref{sec:discussion}
discusses the evolutionary stage of this system, the possible nature of the
gas companion and the implication of our findings. We conclude with a final summary in Section~\ref{sec:summary}.  Throughout
this paper, we adopted a cosmology with $H_0=69.6$, $\Omega_M=0.286$, and
$\Omega_\Lambda=0.714$.

\section{Data Reduction and Analysis} \label{sec:data}

\subsection{Observation and Data Reduction}

HSC J2239+0207 was observed with the NIRCam imager and NIRSpec IFU as part of
the US MIRI GTO program 1205 (PI: George Rieke). The NIRCam results and a
preliminary analysis of the NIRSpec quasar spectrum have been published in
\cite{Stone2023}. This paper focuses on the detailed analysis of the NIRSpec datacube. 

The NIRSpec IFU data of HSC J2239+0207 were obtained on Nov 11, 2022 with the PRISM/CLEAR
disperser-filter combination, providing a $3\arcsec\times3\arcsec$ data cube
over a wavelength range of 0.6--5.3 $\mu$m (corresponding to 825--7300\AA \,at the quasar redshift) at a spectral resolution of 30--300.
A four-point dithering pattern was adopted to mitigate detector artifacts and cosmic rays and the total exposure time on source was about 2.5 hours.

We processed the NIRSpec data with JWST pipeline version 1.12.0 and the CRDS
pipeline parameter reference file jwst\_1088.pmap following the standard steps.
This includes the first stage, \textit{Detector1Pipeline}, to apply
detector-level corrections (e.g., dark current and bias subtraction, persistence
correction, cosmic-ray removal) to the raw data of individual exposures and
produce the un-calibrated 2D spectra; and the second stage,
\textit{Spec2Pipeline}, to assign world coordinate system (WCS) information to
the data, apply flat-field corrections, carry out  flux calibration, and
construct 3D data cubes from the 2D spectra obtained at each dither location.
The third stage, \textit{Spec3Pipeline}, combines the individual data cubes at
different dither positions to produce the final merged data cube with outlier
rejections, drizzling, etc. to remove any additional artifacts and improve
spatial sampling. We found that the default pipeline did not always remove
obvious outliers and added an customized step to mask out the
bad pixels in the 2D spectra manually after a careful visual inspection of individual
spectral slices. After this process, we re-constructed a final IFU cube for science analysis.

\subsection{Spectrum Extraction and Fittings}

\begin{figure*}[htp]
\begin{center}
\includegraphics[width=1.0\hsize]{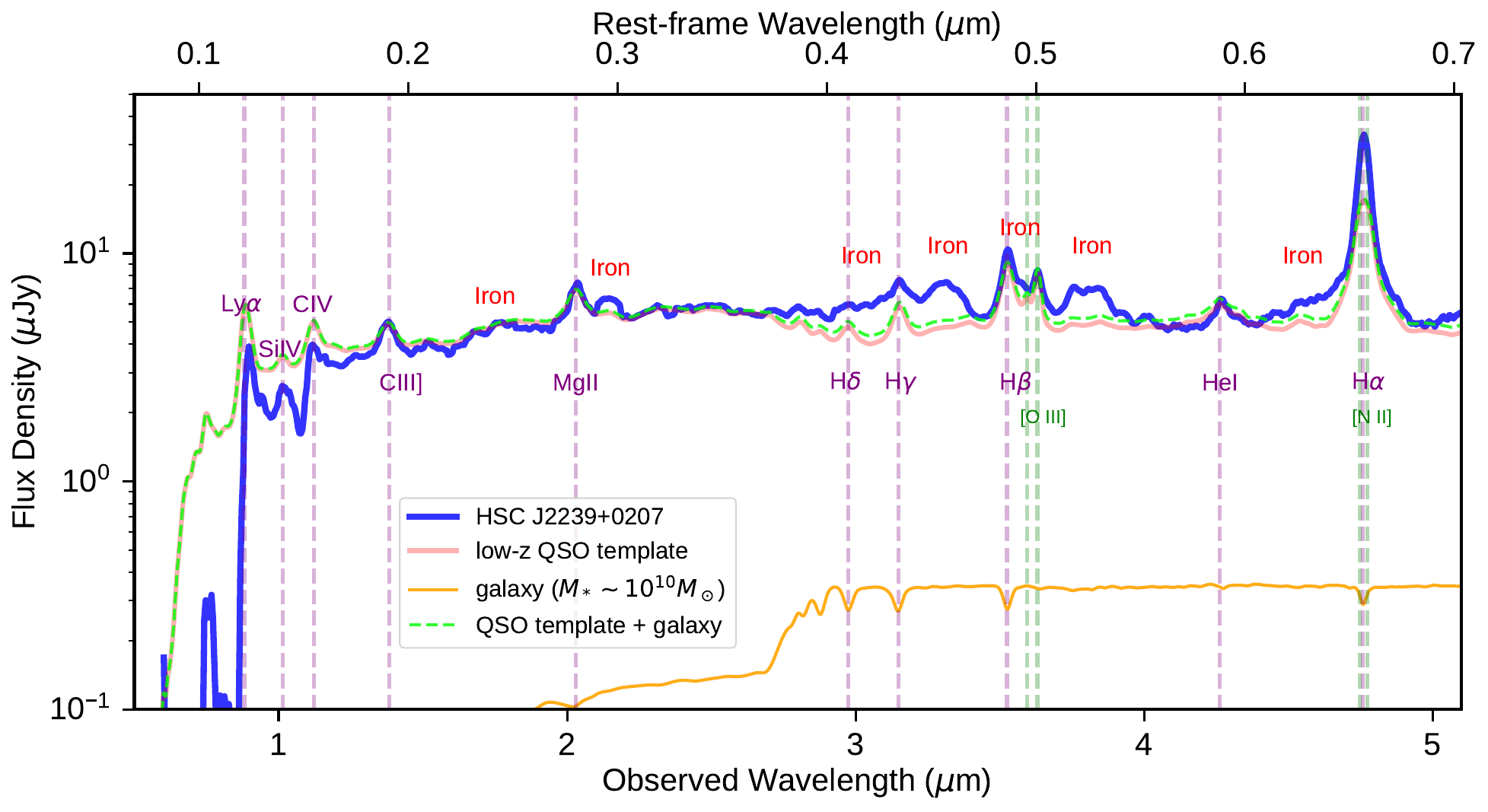}
\caption{The observed nuclear quasar spectrum of HSC J2239+0207 (blue line). A normal quasar template built from SDSS is also plotted in red as a comparison. We have highlighted various spectral features. We have plotted a galaxy template with stellar mass $M_*\sim10^{10} M_\odot$ (orange line) and show the resulting QSO+galaxy spectrum (green dashed line). All the templates have been smoothed to match the NIRSpec prism resolutions.}
\label{fig:quasar_spec}
\end{center}
\end{figure*}

To extract the unresolved nuclear quasar spectrum, we adopt a circular aperture
at a radius of 0.35\arcsec across all wavelengths. For the background
subtraction, we placed the same aperture at random locations of the cube that do
not contain real structures, computed the medium background spectrum, and
subtracted it from the quasar spectrum. We have used WebbPSF \citep{Perrin2014}
to simulate the wavelength-dependent instrument PSF and computed the required
aperture corrections. Our aperture is large enough that the introduced maximum
color offset is less than 8\% across the whole wavelength range. In
Figure~\ref{fig:quasar_spec}, we present the nuclear spectrum of HSC J2239+0207
and compare it to a low-redshift quasar template that has been smoothed to the
NIRSpec prism resolution, which will be discussed in the following section.

\begin{figure*}[htpb]
\begin{center}
\includegraphics[width=1.0\hsize]{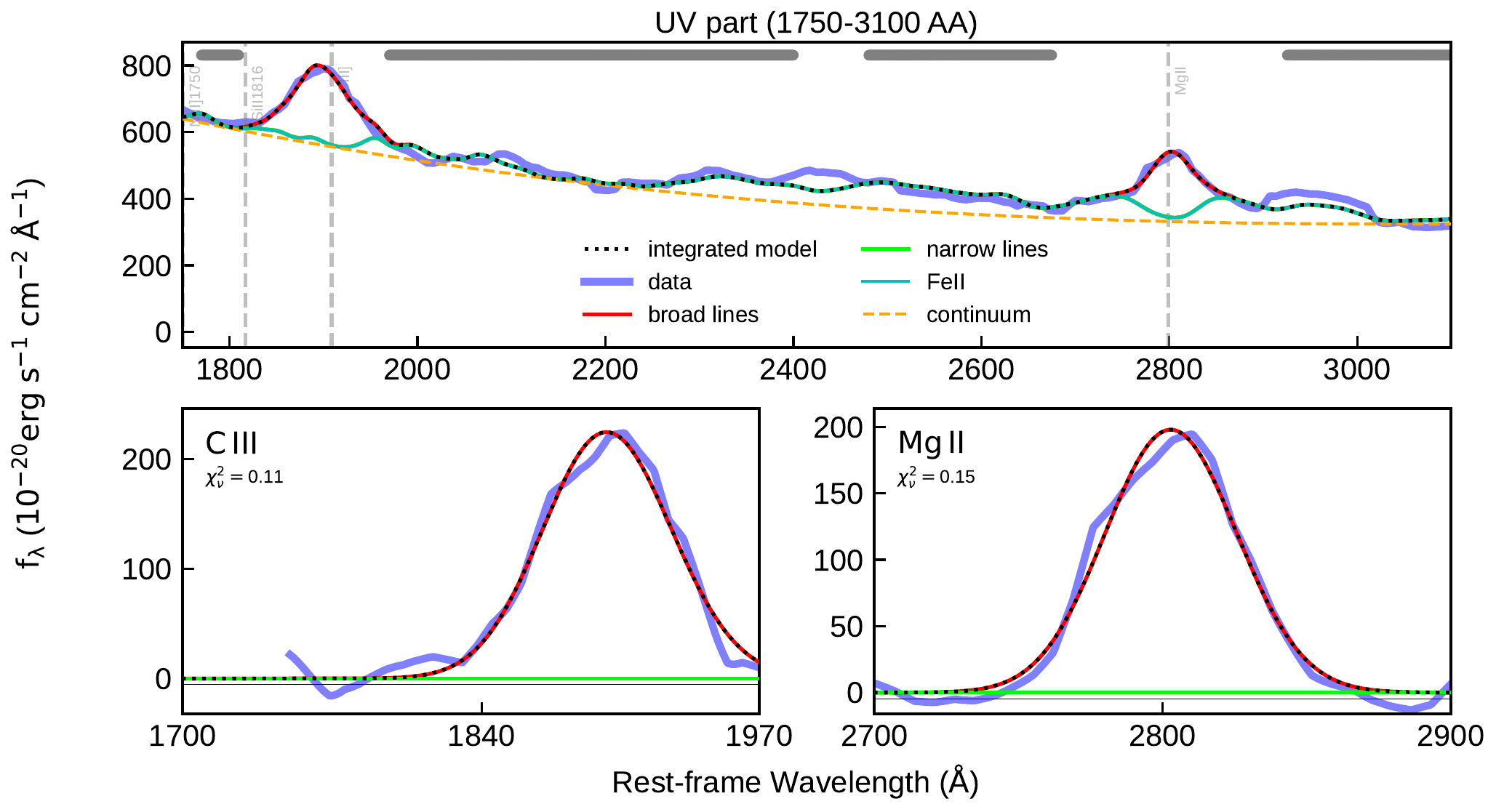}
\includegraphics[width=1.0\hsize]{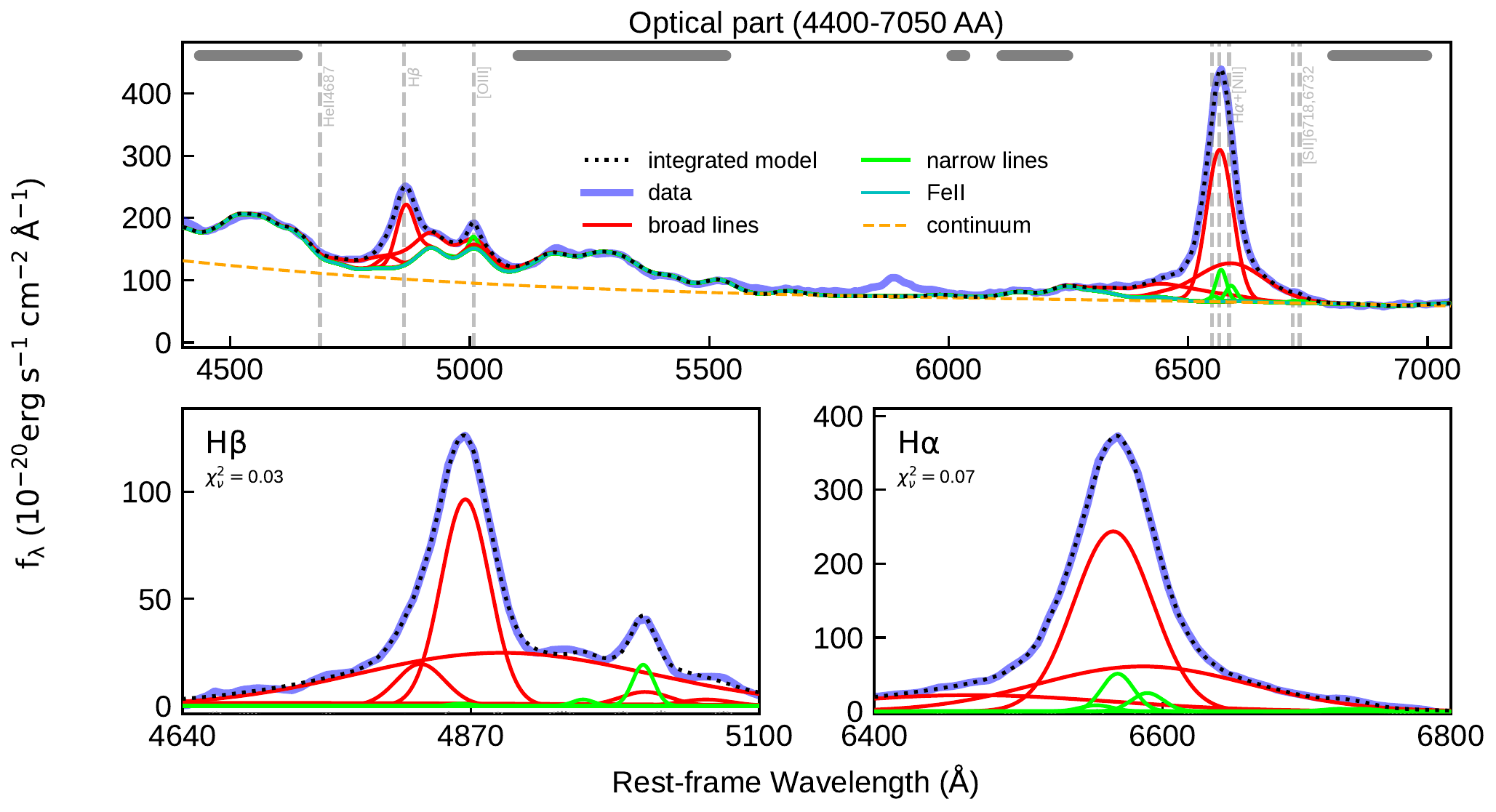}
\caption{PyQSOFit spectral decomposition results of the UV (top) and optical bands (bottom) of HSC J2239+0207 NIRSpec prism spectrum. The integrated model (black dots) is composed of the AGN power-law continuum (dashed orange line), the iron emission model (thin blue line), the broad line components (red lines) and narrow line components (yellow lines). The data is shown in thick light blue line. The wavelengths of some strong nebular emission lines are labeled with vertical gray dashed lines and the gray horizontal bands on the top indicate the wavelength window used to fit the power-law continuum and iron models. }
\label{fig:quasar_spec2}
\end{center}
\end{figure*}

To decompose the quasar spectrum, we have adopted PyQSOFit \citep{QSOfit}.
and conducted the spectral fittings in two wavelength windows: (1) the UV part
from rest-frame 1700~\AA~to 3100~\AA, and (2) the optical part from rest-frame
4400~\AA~to 7050~\AA. First, the continuum was fitted with a three order polynomial function plus
\Feii~templates \citep{Boroson1992, Vestergaard2001} after masking out regions
contributed by emission lines. Then we subtracted the continuum models from the
observed spectra to get the emission-line only spectra.  Emission line fittings were conducted in several different wavelength windows at 6400--6800~\AA ($H\alpha$, \Nii, \Sii), 4640--5100~\AA ($H\beta$, \Oiii, \Heii), 2700--2900~\AA (\Mgii), 1700--1970~\AA (\Ciii) separately. For Hydrogen emission lines ($H\alpha$, $H\beta$), we have adopted three
Gaussians to model the broad-line components with the requirements of the full-width half-maximum (FWHM) $>$ 1000 km/s and one Gaussian to model the
narrow components. For \Ciii~and \Mgii~emission lines where the spectral resolutions are poor, only one broad
Gaussian was used to fit the line profile. For forbidden lines such as \Oiii~and \Sii, two Gaussians were adopted with one for the central component and another for any profile asymmetry, and the profiles of the doublets were tied to be the same with their relative line strength follows the typical theoretical values.  The fitting results are plotted in
Figure~\ref{fig:quasar_spec2}.

As described later, there is an 
elongated emission line gas companion detected
in the datacube. We extracted its signals with an rectangular aperture of
$0.6\arcsec\times1\arcsec$ that covers the spatial distribution,  and we
subtracted the background emission in a similar way to the nuclear quasar
spectrum. { Based on the same WebbPSF model, we found that the  quasar light contamination inside the companion aperture is comparable to or less than the noise level of the companion spectrum with a median relative fraction of 46\%. At H$\alpha$+\Nii~and H$\beta$+\Oiii~ emission line regions where the companion is detected, the peak strength of the expected quasar light contribution is still $\lesssim80$\% of the noise level. As a result, we did not introduce additional corrections. } The emission lines from the companion are not broadened, so a single Gaussian is
adopted to fit  the emission lines, including the unresolved
H$\alpha$+\Nii~complex. 

\section{Quasar Properties} \label{sec:results}

Before diving into the detailed spectral analysis, we first compare the
observed UV-to-optical spectrum of HSC J2239+0207 to quasars at low redshift.
Although the quasar template from \citet{VandenBerk2001} has been widely adopted
as the reference to trace evolution at high-$z$, it does have notable
contamination from the host galaxy (e.g., Ca II and Na I absorptions) and
evidence for dust reddening. We have constructed a new quasar template from
the Sloan Digital Sky Survey (SDSS) Quasar Catalog DR7 \citep{Schneider2010} by
combining the spectra of the  350 quasars with the bluest continuum and
minimum stellar contamination. 
In Figure~\ref{fig:quasar_spec}, we compare the spectrum of HSC J2239+0207 to this quasar
template with the latter smoothed to match the JWST/NIRSpec prism
resolution.

It turned out J2239+0207 is remarkably similar to this low-$z$ template in terms of
the continuum shape for the \Feii-free regions and the profiles of most emission lines. 
There are three notable differences: (1) boosted \Feii~emission
from the UV to the optical; (2) evidence of broad-line-absorptions at
wavelengths shorter in the UV and strong Ly$\alpha$ absorption; (3) asymmetric
profile of the H$\alpha$+\Nii~complex. 

We describe the details of various quasar properties below.

\subsection{Black Hole Mass and Eddington Ratio}

With the decomposition of AGN emission lines (broad/narrow components from
different atoms/ions) and underlying continuum (\Feii~features and AGN featureless continuum), we can get reliable constraints on the mass of
SMBHs from broad Balmer emission lines.

For H$\beta$, we adopt black hole mass estimator in \cite{Vestergaard2006}
\begin{equation}
\frac{M_{\rm BH, H\beta}}{M_\odot} = 10^{6.91}\left(\frac{\rm FWHM(H\beta)}{1000~{\rm km~s}^{-1}}\right)^2  \left(\frac{\lambda L_\lambda(5100{\rm \AA})}{10^{44}~{\rm erg~s}^{-1}}\right)^{0.5}
\label{equ:hbeta-bh-mass}
\end{equation}
We measured the FWHM of the broad H$\beta$
component to be 2992 km/s, after correcting for the instrument broadening (2244.8
km/s) from the apparent $H\beta$ FWHM (3740.2 km/s). The 5100\AA~luminosity is
measured to be $2.04\times10^{45}~{\rm erg/s}$ from the decomposed continuum and the resulting black hole mass
is 3.28$\times10^{8}~M_\odot$.

For H$\alpha$, \cite{Greene2005} provided a BH mass estimator based on the
broad $H\alpha$ emission line alone. 
\begin{equation}
\frac{M_{\rm BH, H\alpha}}{M_\odot} = 10^{6.30}\left(\frac{\rm FWHM(H\alpha)}{1000~{\rm km~s}^{-1}}\right)^{2.06}  \left(\frac{L_{\rm H\alpha}}{10^{42}~{\rm erg~s}^{-1}}\right)^{0.55}
\label{equ:halpha-bh-mass}
\end{equation}
We measured the FWHM of the broad H$\alpha$ component to be $\sim$3284 km/s after taking out the instrument
broadening 1163.9 km/s from the apparent FWHM of 3484.5 km/s. The
H$\alpha$ luminosity is measured to be $1.41\times10^{44}$ erg/s, thus the black
hole mass based on broad H$\alpha$ is $3.51\times10^{8} M_\odot$. 

At shorter wavelengths, the spectral resolution of the prism data is too poor ($\sim6019~{\rm km/s}$ for CIII, $\sim9482~{\rm km/s}$ for Mg II)
to allow useful constraints on the black hole mass. We note that  
\cite{Onoue2019} reported the FWHM of \Mgii~ to be $4670^{+910}_{-700}$ km/s based on ground-based near-IR observations and
estimated a \Mgii-based black hole mass $1.1^{+0.3}_{-0.2}\times10^{9}
M_\odot$, which is about three times larger than the results from H$\alpha$ and
H$\beta$. Similarly, they found a FWHM for  CIV of $4630^{+1040}_{-1260}$ km/s leading to a black hole mass estimate of $6.3^{+2.0}_{-2.5}\times10^{8}
M_\odot$, twice our estimate from H$\alpha$ and
H$\beta$, but consistent within the large error. The differences are not atypical for single-epoch black hole mass estimates, and MgII and CIV specifically may be affected by winds; also the signal to noise of the \citet{Onoue2019} line measurements is low. In addition, our low-resolution spectra show strong \Feii~
emission in the UV part (see the following section). \cite{Onoue2019} might not have been able to constrain this component robustly in their fitting due to the low
S/N of their data, leading to an over-estimation of the line width. It is also possible that \Mgii~method has overestimated the black hole mass of HSC J2230+0207 due to calibration uncertainties and the differences in BLR properties for this particular system.  

The BH mass measurements from H$\alpha$ and H$\beta$ should be relatively reliable given their extensive calibrations; we adopt a final
measurement of $M_{\rm BH}=3.3\times10^8 M_\odot$. The AGN bolometric luminosity
can be estimated from the optical continuum as $L_{\rm bol}=9.26\times L_{5100}$ \citep{Richards2006} and we have $L_{\rm bol}\sim1.9\times10^{46}$ erg/s. As a result, the Eddington ratio of this quasar is 0.44 \footnote{or lower if we use the alternative black hole mass estimates}, confirming the sub-Eddington accretion status of the SMBH in HSC J2239+0207.

\subsection{Proximity Zone}\label{sec:proximity_zone}

The ages of $z\sim6$ quasars can be traced by the sizes of their proximity zones
where the AGN UV radiation ionizes the surrounding intergalactic medium (IGM)
\citep[e.g.,][]{Bolton2007, Eilers2017}. These structures are primarily shaped
by the history of quasar ionizing radiation and the IGM properties,
offering valuable insights into the quasar's activity duration and its local
environment.

\begin{figure}[htp]
\begin{center}
\includegraphics[width=1.0\hsize]{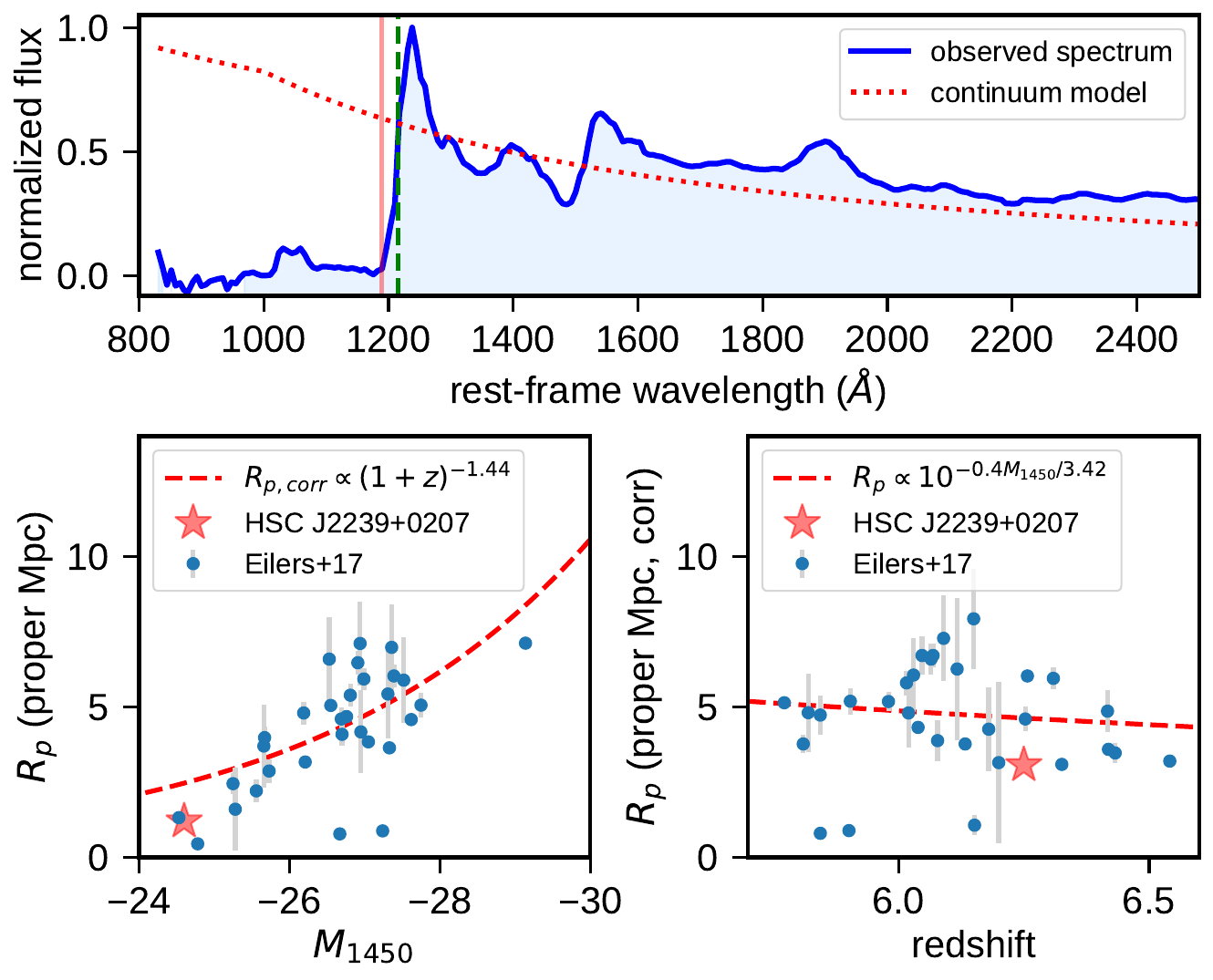}
\caption{Top panel: continuum-normalized spectrum of HSC J2239+0207. The dashed green vertical line denotes the Ly$\alpha$ wavelength (1215.7\AA) and the pink vertical line denotes the wavelength where the quasar flux (blue line) starts to drop below 10\% of the quasar continuum (red dotted line). The wavelength difference between the two ($\delta \lambda$) gives the size of quasar proximity zone. Bottom panels: the location of HSC J2239+0207 on the proximity zone vs AGN luminosity (left) and redshift (right) plane, in comparison to the sample in \cite{Eilers2017}. Dashed lines are the best-fit correlations reported in \cite{Eilers2017}.}
\label{fig:pro_zone}
\end{center}
\end{figure}

To constrain the proximity zone size of HSC J2239+0207, we fitted the quasar continuum at
1241--1285~\AA~and 1310--1380~\AA~by a power-law spectrum with a fixed spectral
index of $\alpha_\lambda=-1.5$ and assumed the continuum at $\lambda<1000$\AA~is
described by a different power law with $\alpha_\lambda=-0.59$ (i.e., a
broken-power model for the AGN continuum; \citealt{Shull2012}). The proximity
zone size is then determined by the wavelength difference between the Ly$\alpha$
wavelength (1215.7~\AA) and where the observed flux drops below 10\% of the
continuum level. (see the top panel of Figure~\ref{fig:pro_zone})

We found a
$\delta\lambda\sim24$~\AA, corresponding to a comoving size of 8.8 Mpc or a
proper size of 1.2 pMpc at $z=6.25$. Based on higher resolution ($R\sim1500$) spectrum and more sophisticated method, \cite{Ishimoto2020} reported a value of 1.65 pMpc for the same quasar. Considering the typical uncertainties of the measured proximity zone sizes \citep[e.g.,][]{Eilers2017} and the different data and methods, these values are in good agreement. As shown in the bottom panels of
Figure~\ref{fig:pro_zone}, HSC J2239+0207 follows the trends defined by other $z\sim6$ bright
quasars \citep{Eilers2017}. A similar conclusion is also reached by \cite{Ishimoto2020}.

From its bolometric luminosity of $1.9\times10^{46}~{\rm erg/s}$, the corresponding ionizing photon production rate $Q_H$ can be estimated to be $\sim3.5\times10^{56}$~photons/s by adopting a typical UV bolometric luminosity ($\sim4$; \citealt{Richards2006}) and ionizing energy $<h\nu>\sim13.6~{\rm eV}$. 
 The quasar lifetime, $t_Q$, can be approximated based on the relationship between
 the size of the proximity zone ($R_p\sim$1.2 Mpc for this quasar) and the
 quasar's ionization history by the following equation
\begin{equation}
t_Q\sim \frac{4\pi R_p^3 n_H}{3Q_H},
\end{equation}
Assuming the surrounding IGM neutral hydrogen number density $n_H\sim10^{-4} {\rm cm}^{-3}$, which is typical for $z\sim6$ \citep{Bolton2007, Eilers2017}, we have a quasar lifetime $t_Q\sim$2 Myr. In real life, the situation is far more complicated, and the estimated value here should be treated as order-of-magnitude \citep[see discussions in e.g.,][]{Eilers2017}. This result reflects the duration of the current active phase during which HSC~J2239+0207 has ionized its surrounding IGM. The proximity zone size, being consistent with the luminosity of the quasar, suggests no evidence of an unusually young quasar or an atypically small ionized region. Instead, the quasar must have been active long enough to ionize its surrounding IGM to a typical extent for its luminosity.

\subsection{Iron Emission}

Iron emission in quasars provide insights into the metallicity and chemical enrichment of the broad-line regions (BLRs), shedding light on the evolutionary stage of these systems.

Following \citet{Dietrich2002}, the UV \Feii~emission was quantified by
integrating the fitted \Feii~ template at rest-frame 2200--3090~\AA. Combining it with the measured flux of the broad \Mgii~line, we get a
flux ratio of \Feii/\Mgii$\sim$3.66. This value is consistent with those in other high-redshift quasars, such as those in the XQR-30 sample \citep{JiangD2024}, indicating that HSC J2239+0207 is metal-rich.

In the optical band, following \citet{Shen2014}, we characterize the \Feii~strength
$R_{\rm\Feii}={\rm EW}_{\rm\Feii}/{\rm EW}_{\rm H\beta}$, where the \Feii~EW is
calculated for the flux within 4434--4684$\AA$ and H$\beta$ is only for the
broad component. The derived $R_{\rm\Feii}\sim1.27$ places this quasar within  Population A on the quasar main sequence \citep[e.g.,][]{Marziani2018}, characterized by relatively high \Feii~emission and moderate H$\beta$ line width (We measure FWHM(H$\beta$)$\sim$3019 km/s for HSC J2239+0207). Compared to the optical \Feii~properties of the eight bright $z\gtrsim6$ quasars observed by the ASPIRE program \citep{Yang2023}, our quasar presents no notable differences despite its lower AGN luminosity, indicating that HSC J2239+0207 shares typical quasar metallicity without evidence of a particularly young or low-metallicity BLR.

\subsection{Winds and Outflows}

\begin{figure}[htbp]
\begin{center}
\includegraphics[width=1.0\hsize]{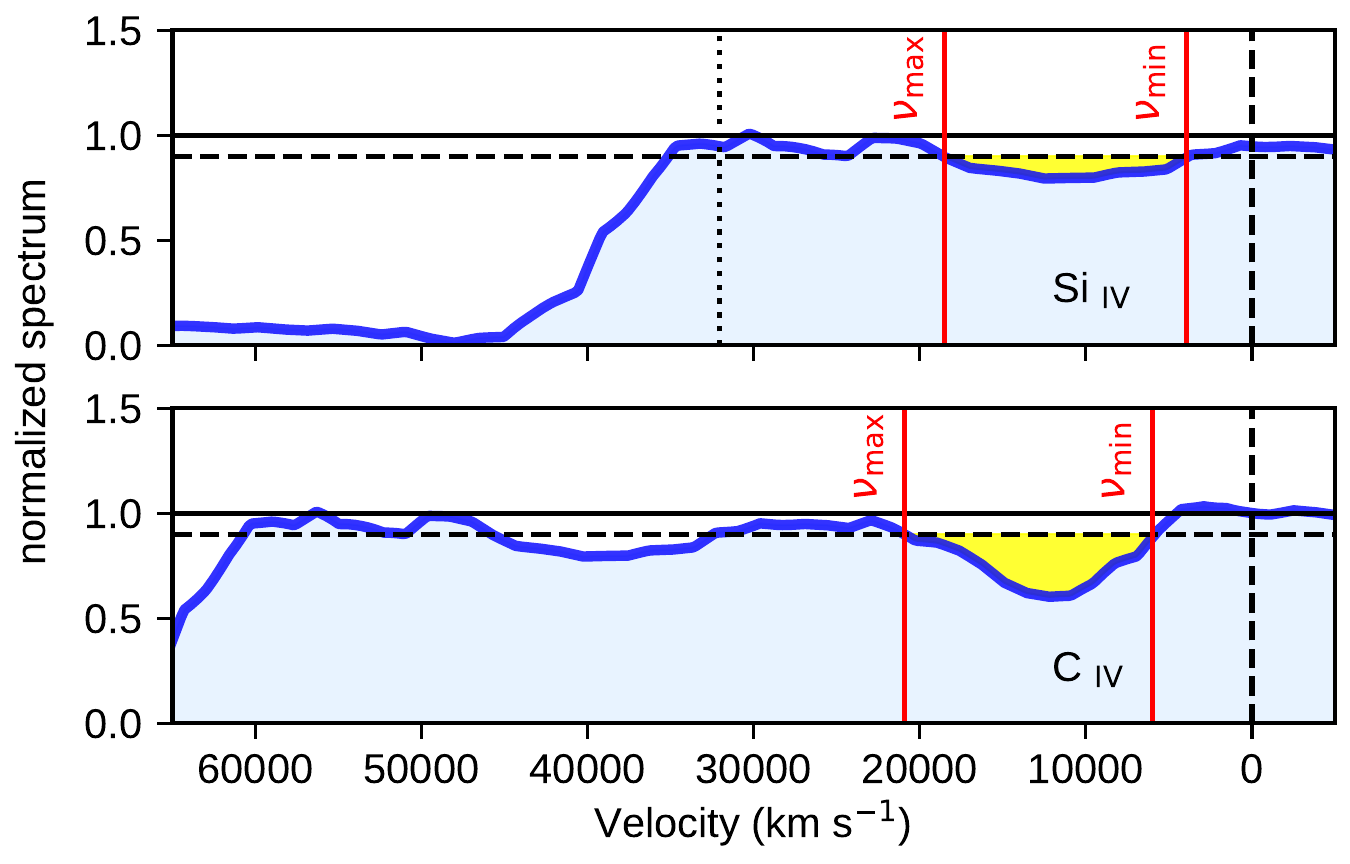}
\caption{Normalized spectrum of HSC J2239+0207 for the BAL feature measurements. The x-axis indicates the velocity relative to the rest-frame wavelength of the ionic species indicated by the label. The solid (dashed) horizontal line represents a normalized flux level of 1.0 (0.9). BAL troughs defined as normalized flux$<$0.9 are highlighted as yellow shaded regions. Vertical red solid lines indicate $\nu_{\rm max}$ and $\nu_{\rm min}$ for each BAL.  }
\label{fig:bal}
\end{center}
\end{figure}

As shown in the comparison to the standard quasar template, HSC J2239+0207 presents
absorption troughs in the rest-frame far-UV, indicating the
presence of broad-absorption line (BAL) features. To quantify the BAL
strengths, we constrain the balnicity index (BI; \citealt{Weymann1991}), a
modified equivalent width of the BAL absorption, following
\citet{Bishetti2023}:
\begin{equation}
BI = \int^{\nu_{\rm lim}}_{0}\left(1-\frac{f(\nu)}{0.9}\right)C d\nu ~,
\end{equation}
where $\nu$ is velocity and $f(\nu)$ is the normalized spectrum, with $C=1$ only when $f(\nu)<0.9$,
otherwise $C=0$. The $f(\nu)<0.9$ limit requires any BAL feature to fall at least 10\% below the continuum. We also identify the minimum and maximum velocity, $\nu_{\rm
min}$ and $\nu_{\rm max}$, for a given transition as the lowest and highest BAL
velocity  where $C=1$. The $BI$ is calculated by normalizing the spectrum of HSC J2239
by the quasar template to match the right wing of \Civ~profile. The results can
be seen in Figure~\ref{fig:bal}. For \Civ, we get $\nu_{\rm min}\sim6000~{\rm
km/s}$, $\nu_{\rm max}\sim20900~{\rm km/s}$ and $BI\sim2700~{\rm km/s}$. A similar absorption trough is evident in the CIV spectrum of \citet{Onoue2019}, with velocity offset and width consistent with the values obtained from the NIRSpec data. For
\Siiv, we have $\nu_{\rm min}\sim3900~{\rm km/s}$, $\nu_{\rm max}\sim18500~{\rm
km/s}$ and $BI\sim1200~{\rm km/s}$. These large velocity ranges, combined with
values of $BI$, indicate the existence of active, high-energy quasar winds in
this system.  

Meanwhile, the profile of the optical H$\alpha$+\Nii~complex shows some
asymmetry with a relatively weak blue-shifted component (see the bottom panel in
Figure~\ref{fig:quasar_spec2}). If this is from the H-alpha component, we estimate
a velocity offset of $\sim$4700 km/s from the blue-shifted broad component, a
value consistent with the minimum BAL velocities. This feature may also be associated with the outflows or winds launched near the quasar in HSC J2239.

\subsection{Dust Attenuation}

From the spectral decomposition, we have derived a Balmer Decrement
(H$\alpha$/H$\beta$)$_{\rm obs}=2.51\pm0.2$ for the quasar broad-line regions, a
value consistent with the ratios observed for the bluest quasars
($\sim2.72\pm0.04$) at low-$z$ \citep[][]{Gaskell2017}. In addition, as seen in
Figure~\ref{fig:quasar_spec}, the quasar
UV-to-optical continuum is in excellent agreement with the blue quasar template once the variations of \Feii~emission are  considered. These observations suggest that the quasar
does not have much dust reddening along the polar direction and indicate the dearth of small grains that cause the UV-optical dust extinction.

\subsection{Host Galaxy}


As described in \citet{Stone2023}, NIRCam images of J2239+0207 show the quasar
host galaxy is very compact with an effective radius of $\sim$1 kpc,
corresponding to $\sim0.17$\arcsec. We have not detected any extended host
signals at the quasar location from the datacube, which is expected given the
limited spatial resolution of NIRSpec IFU (0.1\arcsec$\times$0.1\arcsec~for
every spatial pixel) and the more complicated instrument PSF. The
unresolved quasar spectrum still provides the opportunity to decompose the
integrated light into an AGN and a galaxy component. We do not detect an obvious
host signal with PyQSOFit, likely because of the poor spectral resolution of
the prism does not allow reliable measurements of the weak stellar absorption
lines and the spectral fitting is dominated by broad features such as strong
iron emission and quasar power-law continuum.

The spectrum is better-suited for characterizing broad features, such as the Balmer break. In Figure~\ref{fig:quasar_spec}, we have compared the spectrum of J2239+0207 to
the quasar template with minimal host galaxy contamination. Besides the
obviously stronger Iron emission features and the absorption in the UV, they are
identical, suggesting the host galaxy contribution of J2239+0207 is indeed very
weak, consistent with the an undermassive host galaxy revealed from NIRCam
observations \citep{Stone2023}. Motivated by the stellar mass constrained with
NIRCam from \cite{Stone2023} and the measured host stellar populations of
similar quasars by \cite{Onoue2024}, we have computed a model galaxy spectrum
with a stellar mass of $10^{10}~M_\odot$ at z=6.25 with Flexible Stellar
Population Synthesis (FSPS; \citealt{Conroy2009, Conroy2010}) assuming a
delayed-$\tau$ star formation history with e-folding time of 20 Myr, stellar age
of 200 Myr and attenuation level $A_V=0.1$ and plot it together with the scaled
quasar template.\footnote{We note that the ALMA detection of J2230+0207 yielded an IR SFR of $\sim$450~$M_\odot/{\rm yr}$ \citep{Izumi2019} while the SFR inferred from our model is quite low. However, we do not think this is a problem given the possibilities that (1) the galaxy can be recently quenched while the IR luminosity traces star formation over a timescale of several million years; (2) the young stars associated with very active star formation are likely highly obscured and any detectable stellar emission in the optical is dominated by older stellar population.} As shown by the dotted line in Figure~\ref{fig:quasar_spec}, the contribution of such a host galaxy would be minimal
to the integrated quasar light,
fully consistent with the lack of a stellar component in the PyQSOFit spectral decomposition results.

\section{A Gas Companion}\label{sec:comp}

\begin{figure*}[htp]
\begin{center}
\includegraphics[width=1.0\hsize]{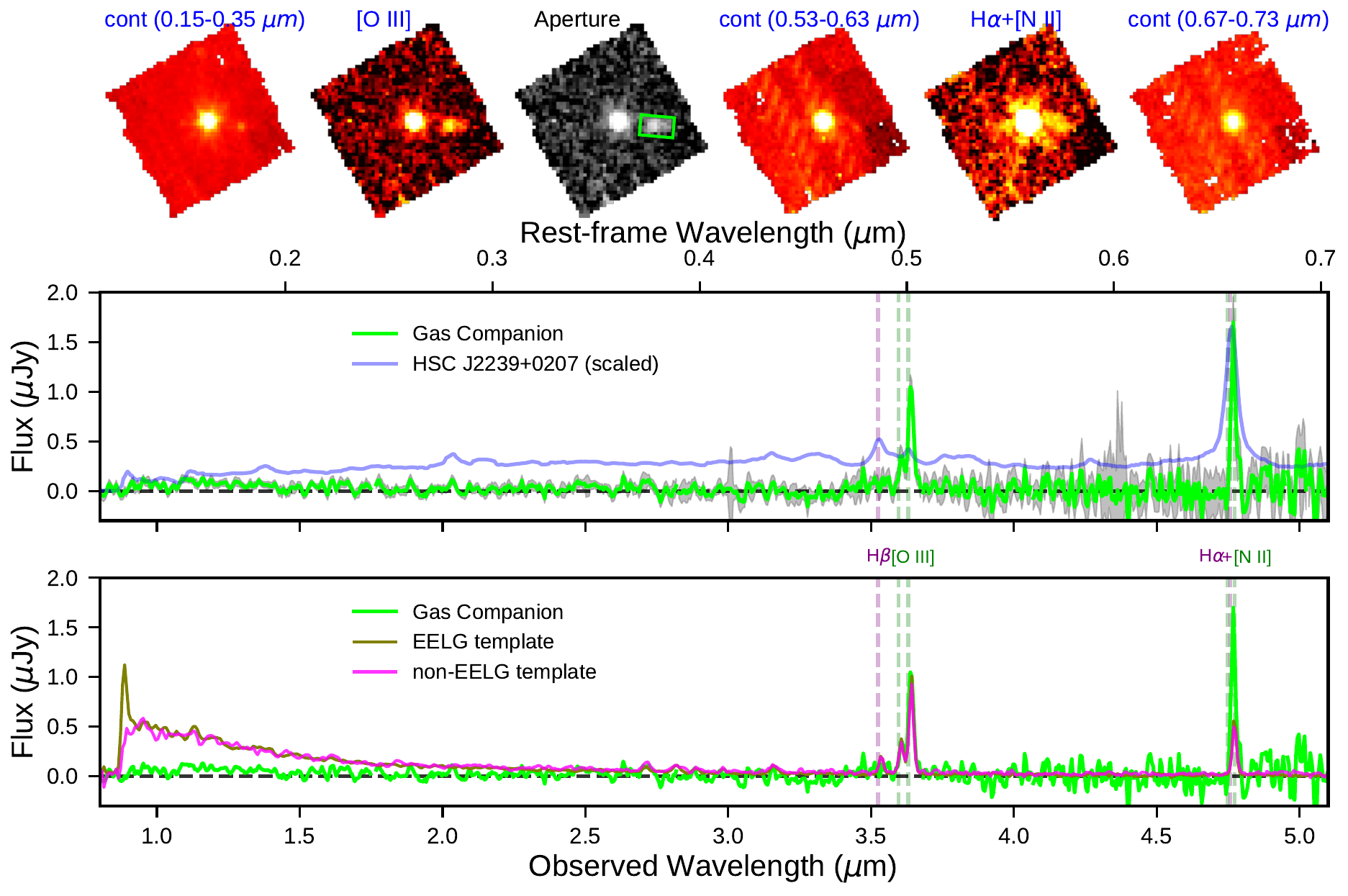}
\caption{Top panel: The IFU re-constructed images of the quasar HSC J2239+0207.
The post-stamps show images at the flux peak of the [OIII] and
H$\alpha$+\Nii~ emission as well as the adjacent emission-line-free continuum. We have highlighted the rectangular aperture used to extract the companion signals in green. Middle panel: The NIRSpec IFU prism spectra of the
emission companion (green line) with the 1$\sigma$ uncertainties (gray shaded regions). We also plot the observed spectrum of quasar HSC J2239+0207 as a comparison (blue line) and labeled the \Oiii, H$\beta$, \Nii, and H$\alpha$ wavelengths at the quasar redshift. Bottom panel: The NIRSpec IFU prism spectra of the companion (green line) compared to two stacked spectra for extreme emission line galaxies (EELGs; dark yellow line) and non-EELGs (magenta line) at $z>5.7$ from \cite{Boyett2024}.}
\label{fig:comp}
\end{center}
\end{figure*}

As shown in Figure~\ref{fig:comp},  near the \Oiii~and
the (unresolved) \Nii+H$\alpha$ wavelengths of the IFU data cube, we have identified an emission-line companion at an
angular separation of $\sim$1 arcsec to the quasar without a significant continuum detection from the UV to the optical. At the quasar redshift, this companion has a projected distance 
of approximately 5 kpc to the quasar. Spatially, it is slightly elongated towards the quasar radial direction with a size
of 0.6$\arcsec\times1\arcsec$, corresponding to a physical extent of
$\sim3\times5$ kpc at the quasar redshift.

On the middle panel of Figure~\ref{fig:comp}, we show the extracted 1-D spectrum
of the companion and compared it to the quasar spectrum. A few lines are
detected for this companion, including the \Oiii$\lambda\lambda$5007, 4959 and
the unresolved H$\alpha$+\Nii~complex. The H$\beta$ emission line is not
convincingly detected. and the strength of H$\alpha$ emission can also not be separated from that of \Nii,~ due to the limited spectral resolution. Compared to the quasar spectrum, all these
lines are shifted towards longer wavelengths. We find that the \Oiii~
emission lines of this gas companion are red-shifted relatively to those from
the quasar by 0.0144$\pm$0.0008 $\mu m$ in the observed frame, corresponding to a velocity offset of
$\sim$1200$\pm$200 km/s at $z=6.25$ (with both the measurement uncertainty and the spectral resolution considered).

By fitting the spectral lines with one single Gaussian function, we find the integrated line strength of the \Oiii~doublet is $I_{[{\rm O{\sevenrm III}},5007]}\sim4.9\times10^{-18}$
erg/cm$^2$/s (11$\sigma$) and $I_{\rm [OIII,4959]}\sim1.6\times10^{-18}$
erg/cm$^2$/s (6$\sigma$). The $H\beta$ emission line is not detected with a 3$\sigma$ upper limit of  $I_{H\beta}\lesssim4.6\times10^{-19}$ erg/cm$^2$/s. The H$\alpha$+\Nii~emission is unresolved in the prism spectrum
with a total line strength about $4.0\times10^{-18}$~erg/cm$^2$/s (6$\sigma$).
Assuming the gas companion does not have extinction in the optical with an
intrinsic ratio $I_{H\alpha}/I_{H\beta}=2.86$, we can put an upper limit of
$H\alpha$ flux, $I_{H\alpha}\lesssim1.3\times10^{-18}$~erg/cm$^2$/s, making the \Nii~emission contribute more than  half of the unresolved flux. Also note
the ratio between the \Nii~doublets $I_{\rm NII,6583}/I_{\rm NII,6549}=3.0$, we
can infer the $I_{\rm NII,6583}\gtrsim2.0\times10^{-18}$ erg/cm$^2$/s and $I_{\rm NII,6549}\gtrsim0.7\times10^{-18}$ erg/cm$^2$/s. The
corresponding flux ratios [N{\sevenrm\,II},6583]/H$\alpha\gtrsim$1.5 and
[O{\sevenrm\,III},5007]/H$\beta$$\gtrsim$10. These values would put the system
into the AGN-dominated area in the classical BPT diagram and can be produced by
either AGN hard radiation (for relatively low ratios) or shocks (for the extreme
ratios) \citep[e.g.,][]{DAgostino2019}

Besides some faint signals in the NIRCam F360M image likely associated with the
\Oiii~emission (see the zoom-in panel in Figure 1 of \citealt{Stone2023}), we do
not detect any emission at the companion location in F210M, F480M and have
put a 3-$\sigma$ upper limit on its stellar mass of
$4\times10^9~M_\odot$, following a similar strategy in \citet{Stone2023}. In addition, ALMA observations of HSC J2239+0207
presented by \citet{Izumi2019} neither reported any [\Cii] flux nor
cold dust continuum at the location of this gas companion.  

We will discuss the nature of this companion and its possible relation to the quasar in Section~\ref{sec:comp-nature}.

\section{Discussion} \label{sec:discussion}

\subsection{HSC J2239+0207 -- A Possibly Fading Quasar}

HSC J2239+0207 provides a unique opportunity to study the evolution of quasars
during the epoch of reionization. Its sub-Eddington accretion rate
($\lambda_{\rm Edd}\sim0.4$)  sets it apart from the majority of bright quasars
at $z\sim6$, which typically accrete near or above the Eddington limit
\citep[e.g.,][]{Inayoshi2020}. This lower accretion rate, combined with its
overmassive SMBH ($M_{\rm BH}\sim3.3$--$3.5\times10^8~M_\odot$) relative to its
compact and undermassive host galaxy \citep{Stone2023}, suggests the system has
undergone an earlier period of rapid SMBH growth and is now entering a less
active phase, {during which the host galaxy may grow later to catch up the black hole mass assembly so that the  local $M_{\rm BH}$-$M_*$ relation can be reached for such systems in the end.}

The estimated lifetime of the quasar stage, on the order of 1 Myr, is inferred from its proximity size of 1.2 proper Mpc (see Section~\ref{sec:proximity_zone}). This highlights the relatively short duration of its recent active phase. This lifetime falls towards the lower end of quasar lifetimes estimated from proximity zones, typically $10^5$ -- $10^7$ years \citep[e.g.,][]{Eilers2017}, suggesting the quasar has not been persistently active throughout its history. Instead, the proximity zone is consistent with the luminosity of the quasar, implying sufficient activtiy to ionize its surrounding IGM but not beyond what is typical for its energy output.

The spectral characteristics of HSC J2239+0207 further support this
interpretation. The strong \Feii~emission and typical proximity zone size
indicate a metal-rich, highly evolved system that has likely been active overall for a
significant duration, presumably in a series of high-activity periods. These properties are in contrast to young quasars or young AGNs, which often show smaller proximity zones \citep{Eilers2020} or significantly reduced
\Feii~emission \citep{Trefoloni2024}. In addition, the presence of high-velocity
BALs in the UV spectrum of HSC J2239+0207, indicative of powerful outflows, along
with minimal dust reddening in the polar direction, suggests that AGN feedback
has regulated the evolution of the SMBH and its environment, potentially clearing the available
accretion material along certain sight lines.

However, while the evidence points toward declining activity, the term
``fading'' must be interpreted cautiously. The sub-Eddington accretion could
reflect a temporary depletion of fuel rather than a permanent transition to
quiescence. Furthermore, the quasar remains luminous and active, with feedback
processes continuing to shape its surroundings, as evidenced by the detection of
a gas companion with high ionization ratios (\Oiii/H$\beta\gtrsim10$) likely driven by
AGN radiation or shocks (see next Section).

\subsection{Nature of the Gas Companion}\label{sec:comp-nature}

The gas companion observed near HSC J2239+0207, located at a projected distance
of $\sim$5 kpc, presents several intriguing characteristics. It exhibits strong
\Oiii~and H$\alpha$+\Nii~emission without notable continuum emission from UV to the optical. The
high \Oiii/H$\beta$ ratio ($\gtrsim10$) places the system in the AGN-dominated
region of the classical BPT diagram, suggesting that the ionization is driven by
AGN hard radiation or shocks rather than star formation. The salient features that guide any discussion are: (1) the ratio of H$\alpha$+\Nii~to H$\beta$ of $\gtrsim$8; (2)
the large extent of the H$\alpha$+\Nii~emission, of order 4 kpc; and (3) the lack of strong UV continuum emission.  Explaining these characteristics is challenging and the true
nature of this companion remains ambiguous, with several possible
interpretations considered below. 

One possibility is that the companion is an isolated high-redshift emission-line
galaxy.  At similar redshifts, a large population of young galaxies are known to show strong nebular emission, particularly
bright \Oiii~lines, at similar redshifts \citep[e.g.,][]{Tang2023, Boyett2024}. In the bottom panel of Figure~\ref{fig:comp}, we have compared the spectral templates of such galaxies from \cite{Boyett2024} to the spectrum of this companion. If we match the \Oiii~line profiles of the companion, these galaxy templates can  explain the lack of stellar continuum emission in the rest-frame optical wavelengths reasonably well, but two notable differences are revealed. First, despite the weaker optical stellar emission, these high-$z$ galaxies typically present strong UV continuum emission from young stars while this companion does not show a significant  detection. Second, the strength of H$\alpha$+\Nii~complex from the companion is atypical compared with  that in the high-$z$ star-forming galaxies. Consequently, this scenario not likely.

Another possibility is that we are seeing a galaxy in the initial stages of merging with the quasar/host galaxy system. The free-fall velocity assuming a quasar mass of $2.9 \times 10^{11}$ M$_\odot$ \citep{Izumi2019} is $\sim$700 km s$^{-1}$, which is consistent with the velocity difference from the quasar if the latter has a small velocity of its own in our direction. In this situation, we expect the ISM of the merging galaxy to be characterized by emission in shocks. This hypothesis provides a straightforward explanation for the strong H$\alpha$+\Nii~emission, since the ISM in merging or interacting galaxies is shocked and the \Nii~lines are significantly enhanced \citep[e.g.,][]{Rich2015,  Mortazavi2019}, up to being comparable to H$\alpha$, i.e., the observed value would be compatible with H$\alpha$/H$\beta $ $\sim$ 3. The largest caution about this hypothesis is that the dynamical mass is quite uncertain due to the effect of the inclination angle on the circum-quasar gas, and that accounting for the large velocity difference is therefore in question. In addition, observations of quasar-galaxy mergers at similar redshifts reveal substantial stellar masses and rest-frame UV emission due to enhanced star formation during the merging process \citep[e.g.,][]{Decarli2019}, which are missing in the companion.

A third scenario is that the gas companion represents tidally disrupted
material from a galaxy passing by. To evaluate this possibility, we estimate the
quasar's halo mass using the empirical relation for galaxies at $z\sim6$ in
\cite{Behroozi2013} and find $M_{\rm halo}\sim4\times10^{11}~M_\odot$ for the
host galaxy's stellar mass $\sim10^{10}~M_\odot$ \citep{Stone2023}. Adopting the
Navarro-Frenk-White (NFW) profile, the halo's virial radius, $R_{\rm vir}$, is
given by
\begin{equation}
    R_{\rm vir}\sim \left(\frac{3 M_{\rm halo}}{4\pi \Delta_{c}\rho_{\rm crit}}\right)^{1/3}, 
\end{equation}
where $\Delta_{c}=2000$ for virialized structures and $\rho_{\rm crit}$ is
the critical density of the Universe at $z\sim6.25$. This calculation yields
$R_{\rm vir}\sim14$~kpc, indicating that the gas companion is within the halo if it
shares the quasar redshift. Assuming the NFW profile, the escape velocity at 5 kpc
is about 200 km/s. However, the redshifted \Oiii~emission of this gas companion
indicates a velocity offset of $\sim$1200 km/s relative to the quasar, far
exceeding this escape velocity. This makes the tidal disruption scenario
unlikely.

A more compelling interpretation is that the gas companion is associated with
AGN-driven outflows. The high velocity, elongated morphology, and alignment
along the quasar's radial direction suggest that it may represent outflowing gas
propelled by AGN feedback. The strong BAL features in the quasar's UV spectrum
indicate powerful winds, and the companion could be a larger-scale manifestation
of these feedback processes. Collimated outflows or shocks may have shaped the
observed structure and sustained the high ionization observed in the gas, as
suggested in e.g., \citet{DAgostino2019}.

The companion could be analogous to low-redshift ``quasar ionization
echoes,'' such as Hanny's Voorwerp \citep{Lintott2009}. In this scenario, the
gas cloud could represent material ionized by the quasar's past activity,
potentially stripped from the host galaxy or another source. However, the
compact and elongated morphology of this companion suggests a more dynamic
origin, such as outflowing gas pushed away by the AGN wind rather than a static
ionization remnant.

In conclusion, the properties of the gaseous quasar companion suggest a variety of interesting explanations, possibly favoring 
an origin tied to AGN-driven feedback. {The real situation can be more complicated with a mixture of various possibilities.} Future deeper JWST NIRCam and high-resolution NIRSpec
observations will be critical for disentangling these scenarios, providing a
definitive understanding of the role of AGN feedback in shaping the environment
of HSC J2239+0207 during the epoch of reionization.

\subsection{Implication on the SMBH Growth and BH-Galaxy Co-evolution}

HSC J2239+0207 offers valuable insights into the complex interplay between SMBH
growth and host galaxy evolution during the epoch of reionization. This quasar
exhibits strong BAL features indicative of significant
gas outflows, yet shows no evidence of dust reddening in its UV-optical
continuum or broad-line Balmer decrement. One possible explanation is that AGN
feedback, through mechanisms such as hard radiation or shocks, has effectively
removed the relatively small dust grains responsible for the UV-optical extinction along the polar direction \citep[e.g.,][]{Laor1993, Tazaki2020}. Under this framework, the gas
companion identified near the quasar may also represent a remnant of earlier
feedback activity.

At longer wavelengths, ALMA observations reveal that the host galaxy of HSC
J2239+0207 is an ultraluminous infrared galaxy (ULIRG) with $L_{\rm IR}\sim
$2.2$\times10^{12} L_\odot$, corresponding to a star formation rate of $\sim$450
$M_\odot {\rm yr}^{-1}$ \citep{Izumi2019}. Compared to the compact far-IR dust
continuum (FWHM$\sim$1.2 kpc$\times$0.7 kpc), its [C II] emission is extended
(FWHM$\sim$4.0 kpc$\times2.6$ kpc) with filamentary structures that trace the
existence of a cold interstellar medium (ISM). Even assuming $\sin i$ = 1, the indicated mass, $6.4 \pm 1.3 \times 10^{10}$ M$_\odot$, significantly exceeds the mass attributed to the host galaxy, implying an exceptionally large mass for this material. Considering the observations
presented in this work, we suggest that although the cold ISM remains available
to sustain star formation, AGN feedback from the quasar has inhibited the growth
of the host galaxy, resulting in an under-massive host stellar component compared to the SMBH
\citep{Stone2023}.

Very recently, \cite{Onoue2024} reported the JWST/NIRSpec fixed slit spectra of
two other HSC quasars (J2236+0032 and J1512+4422), showing post-starburst
signatures in their host galaxies, and suggested that the star formation
activities were recently quenched by the AGN feedback. Although we do not have
direct spectral constraints on the host galaxy stellar population for HSC
J2239+0207 due to the lack of spectral resolution and the fainter host galaxy
signals, HSC J2239+0207 may follow the same evolutionary path as J2236+0032
and J1512+4422, with the AGN feedback strongly regulating the galaxy growth.

\section{Summary} \label{sec:summary}

In this paper, we have presented the NIRSpec/IFU prism observations of HSC
J2239+0207, a low-luminosity quasar at $z\sim6.25$. Despite the low spectral
resolution of the prism data, we have provided various constraints on the
quasar properties and reported a discovery of a gas companion near the quasar
redshift. Our major findings are:
\begin{itemize}
\item The mass of the SMBH in this system is estimated to be (3.3--3.5)$\times10^8~M_\odot$ from
the H$\alpha$ and H$\beta$ broad emission lines and the corresponding Eddington
ratio is 0.44. In contrast to other bright quasars accreting at or above Eddington
limit, HSC J2239+0207 is in a less active phase;

\item We estimate HSC J2239+0207 has a proximity zone size of 1.2 proper Mpc,
consistent with its luminosity and suggesting a recent active phase of on the
order of 1 Myr. This short lifetime places the quasar at the lower end of
proximity zone-derived lifetimes observed in $z\sim6$ quasars, indicating that
this quasar is transitioning out of its peak activity phase; 

\item Consistent with other high-redshift bright quasars, HSC J2239+0207 has strong iron emission and identical spectral properties to its low-$z$ counterparts,  indicating the system is metal-rich and highly evolved and it must have experienced a period (or periods) of active SMBH growth in the past;

\item There are strong BAL features in the UV part of the spectrum that trace strong AGN-driven
winds along the line of sight. However, little dust reddening is revealed from
the broad Balmer decrement and the quasar UV-to-optical continuum. AGN feedback such as hard radiation or shocks can remove the dust grains along the polar direction and explain these observations;

\item We find a gas companion at a distance of $\sim$1 arcsec to the quasar without notable UV-optical continuum emission but a high \Oiii/H$\beta$ ratio and possibly extreme \Nii~ emission line strength, indicating the ionization from an AGN or
shocks. Although we cannot completely rule out other possibilities, considering various constraints, the most favorable explanation for this companion is that this gas cloud was
blown away by the quasar feedback from the host ISM at
some earlier time.
\end{itemize}

Combined with previous NIRCam study of this quasar by \cite{Stone2023}, we have
revealed that HSC J2239+0207 is a sub-Eddington quasar with an under-massive
host galaxy and a gas companion. The AGN feedback traced by the strong BAL
features in the UV is the most reasonable explanation for the slower growth of
the host galaxy despite the rich cold gas reserve revealed by ALMA, as well as the origin
and the high-ionization nature of the gas companion and the lack of dust
reddening of the quasar along the polar direction reported in this work. 

This object provides a unique opportunity to study the subsequent evolution of
bright quasars in the reionization epoch and offers valuable insights into the
interplay between SMBH growth, the host galaxy evolution and the surrounding
environment in the early Universe. Future JWST observations with improved
spectral resolution and sensitivity are desired to enable further study of this
intriguing system.


\begin{acknowledgments}

We thank Dr. Kit Boyett for providing a digital copy of the stacked emission-line galaxy spectra published in \cite{Boyett2024}. J.L., G.R. M.S. and S.A. acknowledge support from the JWST Mid-Infrared
Instrument (MIRI) grant No. 80NSSC18K0555, and the NIRCam science support
contract NAS5-02105, both from NASA Goddard Space Flight Center to the
University of Arizona. The JWST data presented in this paper were obtained from
the Mikulski Archive for Space Telescopes (MAST) at the Space Telescope Science
Institute. { The specific observations analyzed can be accessed via \dataset[doi: 10.17909/dqt1-1e90]{https://doi.org/10.17909/dqt1-1e90}.}

\end{acknowledgments}

\facilities{JWST(NIRCam, NIRSpec)}

\software{Astropy \citep{Astropy1, Astropy2, Astropy3}, Matplotlib \citep{Hunter2007}, NumPy \citep{Harris2020}, SciPy \citep{Virtanen2020}, WebbPSF \citep{Perrin2014}, PyQSOFit \citep{QSOfit}}





\bibliography{bibliography}{}
\bibliographystyle{aasjournal}

\end{document}